\documentclass[12pt]{article}
\usepackage{url} 

\usepackage{amsmath}
\usepackage{graphicx,psfrag,epsf}
\usepackage{enumerate}
\usepackage{natbib}
\usepackage[colorlinks=true, urlcolor=blue, linkcolor=red, citecolor = blue]{hyperref}

\usepackage{tabularx} 
\usepackage{arydshln}
\usepackage{multirow}
\usepackage{amsmath}
\usepackage{amssymb}
\usepackage{arydshln}
\usepackage{amsfonts}
\usepackage[ruled]{algorithm2e}
\usepackage{caption}
\usepackage{subfigure}
\usepackage{subfloat}
\usepackage{ntheorem}
\usepackage{enumerate}
\usepackage{verbatim}
\usepackage{graphicx}
\usepackage{booktabs}
\usepackage{float} 
\usepackage{xr}
\usepackage{xcolor}

\newcounter{xxx}
\newcommand{\eg}{\emph{e.g.}}
\newcommand{\ie}{\emph{i.e.}}

\title{Accelerating Bayesian Phylogenetic Inference via Delayed Acceptance Sequential Monte Carlo with Random Forest Surrogates} 

\author{Wentao Yu$^{1}$, Shijia Wang$^{1\ast}$ \\
{$^{1}$Institute of Mathematical Sciences, ShanghaiTech University,}\\
{ Shanghai, China}\\
{$^\ast$To whom correspondence should be addressed;}\\
{E-mail: wangshj1@shanghaitech.edu.cn.}
}

\begin{document} 

\maketitle

\begin{abstract}
In Bayesian phylogenetics, our goal is to estimate the posterior distribution over phylogenetic trees. Markov chain Monte Carlo methods are widely used to approximate the phylogenetic posterior distributions. For large-scale sequence data, repeated evaluation of the likelihood function incurs a high computational cost. 
In this article, we propose a machine-learning algorithm with over 35 topological and branch-length features to predict the changes in the likelihood function caused by tree moves (\eg,~eSPR, stNNI)  used in standard MCMC approaches. This algorithm is then used to design a delayed acceptance MCMC kernel, which utilized the predicted surrogate function for preliminary rejection, to accelerate tree space searches.
Furthermore, we integrate our proposed MCMC kernel into the sequential Monte Carlo sampler framework. We validate the proposed delayed-acceptance sequential Monte Carlo approach (DA-SMC) on simulation and real data sets. Our delayed acceptance kernel can maintain robust estimation while reduces the number of likelihood evaluations significantly, yielding substantial computational time savings. 
We develop a Python package that is available at \url{https://github.com/wentYu/DAphyloSMC}.

\end{abstract}

\noindent%

\section{Introduction}
A central goal in systems biology is to reconstruct the evolutionary history of biological species (\eg,~DNA sequences). A phylogenetic tree is represented by a tree topology with the associated branch lengths.
Probabilistic evolutionary models are used to describe the stochastic processes of substitutions in modern approaches for phylogenetic inference. Bayesian paradigms are widely used in phylogenetic reconstruction \citep{Lemey2009,drummond2010bayesian,huelsenbeck_mrbayes:_2001,Ronquist2003}. 
The computation of the phylogenetic posterior distribution involves an intractable sum over topologies as well as a high dimensional integral over branch lengths. Markov chain Monte Carlo (MCMC) is a standard method to approximate posterior distributions defined over the space of phylogenetic trees \citep{Rannala1996}. 
Software packages (\eg,~MrBayes \citep{Ronquist2012}, BEAST \citep{suchard2018bayesian, bouckaert2019beast}, and BAli-Phy \citep{Suchard15082006}) have been developed for inferring the phylogenetic posterior using MCMC. \cite{10.1093/sysbio/syab004} propose sophisticated adaptive moves to explore the complex phylogenetic tree space.

Sequential Monte Carlo (SMC) methods \citep{Doucet01, del2006sequential} are good alternatives to MCMC for complex model inference (\eg,~state space models). A growing body of literature applies SMC methods for Bayesian phylogenetic inference. 
\cite{teh08a,gorur09,Bouchard2012Phylogenetic,Gorur2012ScalableSMC,LiangliangWang2015, 10.1093/bioinformatics/btaa867} develop SMC methods that use the Felsenstein pruning recursions to effectively reuse the intermediate targets that are defined over forests over the observed taxa. As the dimensionality of the intermediate targets of these SMC is strictly increasing, the incorporation of phylogenetic MCMC moves is not straightforward. 
To integrate phylogenetic MCMC moves into SMC framework, \cite{wang2018annealed} propose an annealed SMC in the general SMC framework \citep{del2006sequential}. This SMC approach can also be used to estimate the marginal likelihood function, and hence to conduct evolutionary model selection. \cite{dinh2016online, fourment2017effective} design intermediate distributions of SMC to reconstruct phylogenetic tree in an online fashion. 
\cite{everitt2016sequential} apply the reversible jump methods to phylogenetic trees of varying dimensions. \cite{Hajiaghayi2014Efficient, smith2017infectious} apply SMC algorithms to estimate intractable evolutionary models. 

Markov chain Monte Carlo and sequential Monte Carlo both require a large number of samples to ensure algorithmic performance. With the rapid development of sequencing technology, recent phylogenetic studies require analyzing data sets with longer sequences and greater diversity. For large-scale sequencing data, repeatedly evaluating the likelihood function would incur heavy computational burden. \cite{Azouri2021} uses a machine-learning algorithm to
predict the neighboring trees that increase the likelihood, without evaluating the exact likelihood function. 
\cite{zhang2018variational, 10.5555/3495724.3497299, JMLR:v25:22-0348} propose variational approximations to the phylogenetic posterior distributions. In the phylogenetic variational inference, subsplit Bayesian networks are used to represent the tree topology distributions. \cite{xie2023artree} propose a deep auto-regressive model for phylogenetic inference based on graph neural networks.

In this article, we propose a delayed acceptance MCMC (DA-MCMC) \citep{Christen01122005, Joshua2021, Cao03042025} kernel in the Bayesian phylogenetics context, to alleviate the heavy computational burden incurred by repeatedly evaluating the likelihood function. We use a machine learning algorithm to model the functional relationship between an MCMC proposal and the change in log-likelihood function. More specifically, we extract features of nearest neighbor interchange (NNI), extended subtree prune and regraft (eSPR) and multiplicative branch length proposals, then use random forest (RF) to select important features and to predict the change in the log‑likelihood. The resulting trained RF function is used to predict the likelihood ratio. In addition, we embed our DA-MCMC kernel into the sequential Monte Carlo framework to improve the quality of posterior inference. 

The rest of this article is organized as follows. In Section \ref{sec:bac}, we show some background of Bayesian phylogenetics and the notations used throughout. In Section \ref{sec:met}, we introduce our proposed approach. We demonstrate our method using numerical studies in Section \ref{sec:sim} and \ref{sec:real}. We conclude in Section \ref{sec: conclusion}.

\section{Background and Notation}
\label{sec:bac}

We let $D$ denote the observed sequence alignment,  and let $X$ denote a set of observed taxa. We use $t$ to denote a phylogenetic $X$-tree, which is composed of a tree topology  $\tau$ and a set of branch lengths $v$. We use $\theta$ to denote the unknown parameters in a probabilistic evolutionary model (\eg,~the general time reversible (GTR) model) that describes the process of substitutions. Given a tree $t$ and an evolutionary parameter $\theta$, we can compute the likelihood of data, denoted by $p(D|t,\theta)$, using Felsenstein's pruning algorithm \citep{felsenstein1973maximum,felsenstein1981evolutionary}. 

In Bayesian phylogenetic inference, we assign prior distributions to $t$ and $\theta$, denoted by $p(\theta, t)$. It is often assumed that the priors for $\theta$ and $t$ are independent (\ie,~$p(t,\theta)= p(\theta)\cdot p(t)$). 
We are interested in estimating the posterior distribution of $(t, \theta)$,
\begin{equation}
\label{eq:1}
\pi(\theta,t)=\frac{p(\theta)p(t)p(D|\theta,t)}{\int_{\theta}\int_{t}p(\theta)p(t)p(D|\theta,t)d\theta dt}.
\end{equation}
The integral in the denominator involves summing up over all possible tree topologies, which is numerically prohibitive. 

Markov Chain Monte Carlo (MCMC) algorithms, more specifically Metropolis Hastings (MH) algorithms, are standard approaches in Bayesian phylogenetic inference. For brevity, we let $x=(\theta,t)$, and use $\gamma(\cdot)$ to denote the unnormalized posterior density. In the phylogenetic MCMC approach, we iterate between the following three steps until convergence is achieved.  
At $(r + 1)$-th MCMC iteration, we repeat the following procedure.
\begin{itemize}
    \item Propose a new tree and/or new evolutionary parameters, $x_{r+1}^*\sim q(x_r,x_{r+1}^*)$. 
For example, using a nearest neighbor interchange (NNI) on tree topology, and/or a random walking proposal on branch lengths and/or evolutionary parameters.
    \item Compute the Metropolis-Hastings ratio based on $\gamma$:
\[\alpha(x_{r},x_{r+1}^*)=\min\{1,\frac{\gamma(x_{r+1}^*)q(x_{r+1}^*,x_r)}{\gamma(x_r)q(x_r,x_{r+1}^*)}\}\]
    \item Accept the proposed $x^*$ with probability $\alpha(x_{r},x_{r+1}^*)$.
\end{itemize}
Once the MCMC has converged, the proposed samples of $x^*$ are exactly sampled from the posterior distribution.   

We provide a summary of notations used in the paper in Supplementary Appendix S.5. 

\section{Methods}
\label{sec:met}
In Bayesian phylogenetics, the main computational cost of MCMC-based approaches is the evaluation of the likelihood function. To improve the efficiency of the algorithm, we propose a delayed acceptance MCMC kernel in Section \ref{sec: daphylo}. In Section \ref{sec: mlphylo}, we propose a machine learning approach to model the change in the log-likelihood function caused by an MCMC proposal (\eg,~eSPR move). In Section \ref{sec: dasmc},  we combine the proposed DA-MCMC kernel with SMC to improve algorithmic performance.  In Section \ref{sec: train},  we introduce an efficient training data collection procedure.  

\subsection{A delayed acceptance MCMC kernel for Bayesian phylogenetics}
\label{sec: daphylo}

For large-scale sequencing data sets, the computational cost of calculating the likelihood function dominates the rest in the MCMC steps. In this article, we propose a delayed acceptance MCMC kernel to reduce the computational burden incurred by likelihood evaluation of Bayesian phylogenetics. The proposed DA-MCMC kernel admits the following form
\begin{equation}
\label{eq: 2}
    \hat{\alpha}(x, x^\star) = \min\big\{1, \mathbf{1}(\Delta_{x^\star}>\kappa)\frac{p(x^\star)q(x|x^\star)\min\{p(D|x^\star), f(x^\star)\}}{p(x)q(x^\star|x)\min\{p(D|x), f(x)\}}   \big\},
\end{equation}
where $x$ denotes the current state of parameters, $x^\star$ denotes a proposed state, $\Delta_{x^\star}$ denotes the surrogate change in log-likelihood function $\Delta_{x^\star} = \log{f(x^\star)} - \log{p(D|x)}$, $\kappa$ is a negative threshold value pre-specified by users, which represents the tolerable lower bound of predicted log-likelihood change. 

In practice, this DA kernel can be implemented in three steps. 

(1) Firstly, we can compare the surrogate change in log-likelihood function after a state proposal $\Delta_{x^\star}$ to the threshold $\kappa$. If $\Delta_{x^\star}  < \kappa$, we reject $x^\star$ without evaluating the ratio of likelihood functions, the priors and the proposals, since under this circumstance we believe that the negative change of likelihood function dominates the acceptance rate. 

(2) Secondly, if $x^\star$ is not rejected, we compute the ratio of priors and proposals, and accept $x^\star$ with probability 
\begin{equation}
\label{eq: 3}
    \tilde{\alpha}(x, x^\star) = \min\big\{1, \frac{p(x^\star)q(x|x^\star)f(x^\star)}{p(x)q(x^\star|x)\min\{p(D|x), f(x)\}}   \big\}.
\end{equation}
This step does not involve the computation of new likelihood $p(D|x^\star)$, while it only involves evaluating the ratio of priors and proposals. The cost is much cheaper than computing the true likelihood of a phylogenetic tree. If a proposed state is accepted, we then turn to the last step. 

(3) Finally, we accept $x^\star$ with probability
\begin{equation}
\label{eq: 4}
    \bar{\alpha}(x,x^*)=\min\big\{1, \frac{p(x^\star)q(x|x^\star)\min\{p(D|x^\star),f(x^\star)\}}{p(x)q(x^\star|x)\min\{p(D|x), f(x)\}}\big\}.
\end{equation}
If $x^\star$ passes all three examinations, we can tell it is accepted by the overall acceptance ratio $ \hat{\alpha}(x, x^\star)$ in Eq. (\ref{eq: 2}).

In addition, we introduce a likelihood value adjustment parameter $\delta$, motivated by providing a method for manually adjusting the predicted likelihood value to balance the computational efficiency and the posterior accuracy. $\delta$ is directly added to the surrogate log-likelihood change $\Delta_{x^\star}$, and then the adjusted $\Delta_{x^\star}$ will be used in all steps of the DA kernel. From Eqs. (\ref{eq: 3}) and (\ref{eq: 4}), it can be inferred that a larger $\delta$ will lead to few rejections at the early rejection steps, which will result in higher posterior accuracy. However, it will also be more time-consuming due to the calculation of more true positives. When $\delta$ is extremely large, DA-MCMC will degrade into a standard MCMC. We will discuss the setting and detailed effects of $\delta$ in Section \ref{sec: train} and Section \ref{sec: deltasensitivity}.

Algorithm \ref{algo:OneDA} summarizes one MCMC iteration of our DA-MCMC for Bayesian phylogenetic inference. 
In the second step, if a uniform random number $u >  \tilde{\alpha}(x, x^\star) $, then $u$ is greater than the acceptance probability displayed in Eq. (\ref{eq: 4}). Hence, if the surrogate likelihood function is very close to the true likelihood function, we can avoid the likelihood computation in most rejection cases. In this DA-MCMC algorithm, all accepted samples will be required to evaluate the true likelihood, thus the previous likelihood $p(D|x)$ in the denominator of Eqs. (\ref{eq: 2}), (\ref{eq: 3}) and  (\ref{eq: 4}) is always available.

\noindent\textbf{Assumption 1:} The intersection of the supports of $f$ and $\pi$ has positive measure.

\noindent\textbf{Proposition 1:} If the surrogate likelihood function $f$ satisfies Assumption 1, the proposed DA-MCMC admits the following distribution as invariant
\begin{equation}
\label{eq: 5}
    \hat{\pi}(x) = \frac{\mathbf{1}(\Delta_{x}>\kappa)p(x)\min\{p(D|x), f(x)\}} {\int \mathbf{1}(\Delta_{x}>\kappa)p(x)\min\{p(D|x), f(x)\}dx}.
\end{equation}

The detailed balance condition of the DA-MCMC kernel can be verified using Eqs.~(\ref{eq: 2}) and (\ref{eq: 5}). We refer readers to Supplementary Appendix S.1 for a more detailed proof. 

Proposition 1 shows the resulting posterior distribution of the DA-MCMC algorithm, which is different from the standard phylogenetic posterior distribution. The difference between the two posterior distributions originates from two sources. The first is the initial check of surrogate change in log-likelihood function $\mathbf{1}(\Delta_{x}>\kappa)$, and the second is the evaluation of proposed sample $x^*$ using \(\tilde{\alpha}(x, x^\star)\). 

If our selected surrogate change of phylogenetic likelihood function can perfectly model the real change and if the threshold $\kappa$ is properly selected, then \(\hat{\pi}(x)\) will be equal to the true posterior distribution. In the case that the surrogate change underestimates the true value, we may incorrectly reject samples that should be accepted with some probability, and this probability is a function of the surrogate change. Since we reject many samples for which the likelihood is supposed to be estimated, the computing speed of the algorithm is much faster. Conversely, if the surrogate change overestimates the true value, then we reject fewer samples using the pseudo acceptance probability and the computing speed is slower. 
For each accepted sample, we compute its likelihood value. Hence, we may falsely reject samples (Type II error). 
From the theoretical analysis of our acceptance ratio, we may also falsely accept samples (Type I error) under specific circumstances. In actual experiments, we find the number of false acceptances is far smaller than that of false rejections. The frequency of both types of errors is synchronized, and can be controlled at a low level by tuning parameters. We carry out an algorithm to control the proportion of Type II error in Section \ref{sec: train}.

\begin{algorithm}[h!]
\SetAlgoLined
\SetKwFunction{IL}{InitializeDistance}
\SetKwFunction{PL}{PropagateInsertion}
\SetKwFunction{MIN}{Min}
\SetKwFunction{MX}{Max}
\SetKwFunction{TOP}{Top}
\SetKwFunction{Push}{Push}
\SetKwFunction{Pop}{Pop}
\SetKwFunction{Append}{Append}
\SetKwData{Queue}{Queue}

\KwIn{The current parameters $x_r$, $r>0$, surrogate abandoned lower bound $\kappa$, likelihood tuning bias $\delta$.}
\KwOut{Parameters of the next iteration $x_{r+1}$}
\textit{Propose} a new tree and/or new evolutionary parameters, $x_{r}^*\sim q(x_{r-1},x_{r}^*)$. 
For example, using a nearest neighbor interchange (NNI) on tree topology, and/or a random walking proposal on branch lengths and/or evolutionary parameters.
\\\textit{Collect} the set of features of this proposal.
\\\textit{Obtain} the predicting result of the features $\Delta_{x^*_{r+1}}$ by RF.
\\\uIf {$\Delta_{x^*_{r+1}}<\kappa$}{
           \textit{Reject} $x_{r+1}^*$ directly, and set $x_{r+1}=x_{r}$
           \\\KwRet{$x_{r+1}$}\;
     }\Else{
         $\Delta_{x_{r+1}^*}=\Delta_{x_{r+1}^*}+\delta$ 
         \\$f(x_{r+1}^*)=e^{\Delta_{x_{r+1}^*}}p(D|x_{r})$
         \\\textit{Compute} the acceptance ratio $\tilde{\alpha}(x, x^\star) = \min\big\{1, \frac{p(x^\star)q(x|x^\star)f(x^\star)}{p(x)q(x^\star|x)\min\{p(D|x), f(x)\}}\big\}$
         \\\textit{Simulate} $u\sim \mathbf{U}(0,1)$
         \\\uIf{$u\leq\tilde{\alpha}$}{
            \textit{Compute} the second acceptance ratio $\bar{\alpha}(x, x^\star)=\min\big\{1, \frac{p(x^\star)q(x|x^\star)\min\{p(D|x^\star),f(x^\star)\}}{p(x)q(x^\star|x)\min\{p(D|x), f(x)\}}\big\}$
            \\\uIf{$u\leq\bar{\alpha}$}{
                $x_{r+1}=x_{r+1}^*$\;
            }\Else{
                $x_{r+1}=x_{r}$\;
                }
          } \Else {
                $x_{r+1}=x_{r}$\;
          }
     }
 \KwRet{$x_{r+1}$}\;
 \caption{One step of the delayed acceptance MCMC}
 \label{algo:OneDA}
\end{algorithm}

\subsection{A machine learning approach for surrogate likelihood function}
\label{sec: mlphylo}

Machine learning methods cannot directly be applied to model the relationship between a phylogenetic tree and the likelihood function, as the phylogenetic tree topology is defined on a discrete state space. In this article, we propose to use machine learning, more specifically random forest (RF) algorithm, to model the change in log-likelihood function caused by a tree proposal and the continuous features that are used to represent the tree propagation step. 

\begin{figure}[ht]
    \centering
    \includegraphics[width=0.8\linewidth]{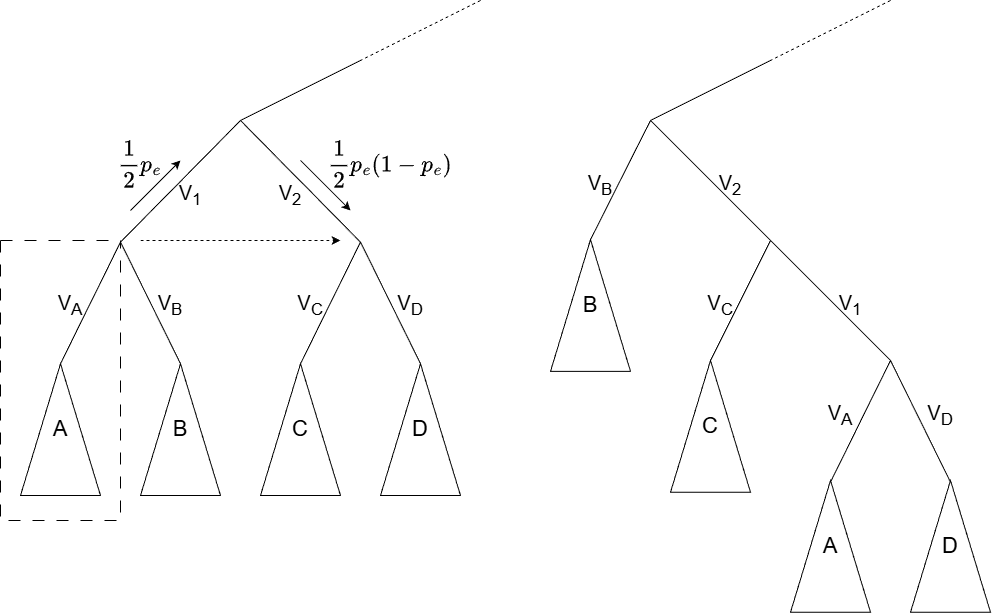}
    \caption{An example of extending Subtree Pruning and Regrafting (eSPR) move. Step 1: Randomly select a subtree as the pruning part. Step 2: Randomly choose a direction to search for the regrafting point. Step 3: With probability $p_e$, extend the regrafting point to the next adjacency point. Step 4: Repeat Step 2, 3 until extending process stop in Step 3 and nail down the regrafting point. Step 5: Prune the pruning subtree and regraft it onto the regrafting point.}
    \label{fig:eSPR}
\end{figure}

We mainly use four types of tree proposals, including eSPR, stNNI, multiplier and global multiplier \citep{Lakner01022008}, among which eSPR and stNNI focus on both topological and branch length moves and the other two only modify the branch lengths. These four tree proposals operate as follows. Note that both eSPR and stNNI may change several branch length after the topological move, utilizing the same method as Multiplier.
\begin{itemize}
    \item eSPR: This move randomly selects a subtree as the pruning part, and then it chooses another branch with a descending probability according to the distance away from the pruning point. The chosen position is the regrafting point where the pruned subtree is reattached. Figure \ref{fig:eSPR} provides an illustrative example of an eSPR move \citep{Lakner01022008}.
    \item stNNI: It is a simpler proposal that randomly selects an intermediate branch and then randomly chooses two of its four neighboring subtrees for swapping. When a certain neighboring subtree is selected, there is a $1/3$ probability of choosing the subtree on the same side, leading to no topology rearrangement. 
    \item Multiplier: This move alters one branch length by multiplying it with $m=e^{\lambda(u-0.5)}$, where $\lambda$ is a parameter and $u$ is a random variable with uniform distribution. 
    \item global Multiplier: It applies Multipliers on all branches within one proposal.
\end{itemize}
We extract continuous features of these tree proposals. These features should include as much information of a tree proposal as possible. We mainly consider eSPR as a standard feature-extraction target, since other MCMC tree proposals can be regarded as special cases of eSPR and are also suitable for these features. Table \ref{tab:feature_tab} displays all $35$ features extracted for an eSPR move. Some features are extracted from the tree before the eSPR operation, and others are based on the pruning and regrafting operations, representing different parts or procedures in the eSPR move. The features we propose are more comprehensive both globally and partially than \cite{Azouri2021}. They can be categorized into global features, partial features, rearrangement features, and heuristic features based on their representation. 

The global features capture the branch lengths information of the entire tree. The partial features capture the topology and branch information of four subtrees divided by the pruning edge and the regrafting edge. The rearrangement features capture the eSPR movement-related information, concentrating on the path from pruning edge to regrafting edge, along with the influenced subtrees on the path.
Heuristic features are indicators for some significant phenomena which may greatly influence the likelihood. In practice, heuristic features are usually some combinations of other types of features, thus may create redundancy. 

Here we explain several important features. The feature `Total branch lengths influenced' is a rearrangement feature, which considers all subtrees on the path from the pruning edge to the regrafting edge, including the path itself and the two subtrees of the pruning and regrafting edges, and sums up all branch lengths of them. The feature `Number of species in subtrees 1' is a partial feature, which counts the number of leaf nodes in subtree no.$1$, namely the pruning subtree. The feature `Long branch attraction risk' is a heuristic feature, which is realized as the product of feature `Number of species in subtrees 1' and the value of feature `Regrafting branch length', which indicates the risk of moving a large subtree onto a long branch and causing long branch attraction (LBA). More detailed descriptions of the selected features are summarized and illustrated in Supplementary Appendix S.2 and S.3.

It is notable that under complex models like K2P or GTR, some evolutionary parameter proposals are introduced, like the base frequency Dirichlet proposal and the evolution rate Dirichlet proposal. These proposals correspond to Euclidean parameter space rather than complex tree space, thus adding their corresponding features is also very intuitive, which uses the parameter values before and after the proposal as features directly. In addition, considering that the proportion of evolutionary parameter proposals in the proposal set is not significant, bringing them into the delayed acceptance framework provides limited help in accelerating the process.

\begin{table}
    \centering
      \caption{Selected features of eSPR phylogenetic tree move.}
      {\footnotesize 
    \begin{tabular}{lccl}
          \hline
          No&Feature name&Representation&Explaination\\
          \hline
          1&Move mode&Heuristic &\begin{tabular}{l}An explicit indicator whether\\regrafting can happen\end{tabular}\\
          2&Total branch lengths&Global&\begin{tabular}{l}The sum of branch lengths \\in the starting tree\end{tabular}\\
          3&Longest branch&Global&\begin{tabular}{l}The maximum branch length \\in the starting tree\end{tabular}\\
          4&Variance of branch lengths&Global&\begin{tabular}{l}The variance of branch lengths\\in the starting tree\end{tabular}\\
          5&Prunning branch length&Rearrangement&\begin{tabular}{l}The length of the prunning branch\end{tabular}\\
          6&Regrafting branch length&Rearrangement&\begin{tabular}{l}The length of the regrafting branch\end{tabular}\\
          7&Prunning and regrafting branch length ratio&Rearrangement&\begin{tabular}{l}The ratio of the prunning \\branch length and the \\regrafting branch length\end{tabular}\\
          8&Number of species influenced&Rearrangement&\begin{tabular}{l}The number of leaves of all \\subtrees along the path\end{tabular}\\
          9&Total branch lengths influenced&Rearrangement&\begin{tabular}{l}The total length of subtrees along the \\path from the prunning node \\to the regrafting node\end{tabular}\\
          10&Longest branch influenced&Rearrangement&\begin{tabular}{l}The maximum branch length\\ of subtrees along the path\end{tabular}\\
          11&Variance of branch lengths influenced&Rearrangement&\begin{tabular}{l}The variance of branch length\\ of subtrees along the path\end{tabular}\\
          12&Topology distance from the pruned node&Rearrangement&\begin{tabular}{l}The number of nodes on the \\path from the pruned node \\to the regrafted node\end{tabular}\\
          13&Branch length distance from the pruned node&Rearrangement&\begin{tabular}{l}The sum of branch length on the\\ path from the pruned node \\to the regrafted node\end{tabular}\\
          14&Longest branch from the pruned node&Rearrangement&\begin{tabular}{l}The maximum branch length on the\\ path from the pruned node \\to the regrafted node\end{tabular}\\
          15&Variance of branch lengths from the pruned node&Rearrangement&\begin{tabular}{l}The variance of branch length on the\\ path from the pruned node \\to the regrafted node\end{tabular}\\
          16-19&Number of species in subtrees 1-4&Partial&\begin{tabular}{l}SubTree 1-4 reference in Fig X\end{tabular}\\
          20-23&Total branch lengths in subtrees 1-4&Partial&The total length of each subtree\\
          24&Subtree 1 and subtree 3 branch length ratio&Heuristic&\begin{tabular}{l}The ratio of pruning and \\regrafting subtree size, $-1$ \\when the regrafting not exist\end{tabular}\\
          25-28&Longest branch of subtrees 1-4&Partial&\begin{tabular}{l}The maximum branch length \\of each subtree\end{tabular} \\
          29-32&Variance of branch lengths of subtrees 1-4&Partial&\begin{tabular}{l}The variance of branch length \\of each subtree\end{tabular} \\
          33&Long branch attraction risk&Heuristic&\begin{tabular}{l}The pruning tree size $*$ the \\regrafting branch length, indicating \\the risk of moving a large subtree\\ onto a long branch\end{tabular}\\
          34&Pruned and regrafted branch length sum&Rearrangement&\begin{tabular}{l}Pruning branch length\\ $+$ regrafting branch length after \\applying multiplier on branches\end{tabular}\\
          35&Pruned and regrafted branch length product&Rearrangement&\begin{tabular}{l}Pruning branch length\\ $*$ regrafting branch length after \\applying multiplier on branches\end{tabular}\\
    \end{tabular}
    }
    \label{tab:feature_tab}
\end{table}

In this article, we use random forests (RFs, \citealp{Breiman2001Random}) to predict the change in log-likelihood function using the extracted phylogenetic features. There exist several advantages of using RFs over other classical machine learning approaches. Firstly, the foremost strength is their robustness to uninformative features, allowing the inclusion of high-dimensional features without pre-selection, which is suitable for exploration without sufficient experience. Secondly, RFs inherently possess the ability to quantify the importance of features, enabling feedback of phylogenetic feature selection. Thirdly, RF is robust to large and nonlinear sample data, and has fast prediction speed among machine learning approaches.

An RF aggregates an ensemble of $B$ regression trees, each trained on a bootstrap sample of the dataset. During the tree construction, binary splits at internal nodes partition the feature space using rules like $ X^{(j)} \leq s$, where the chosen feature $X^{(j)}$ and split point $s$ are optimized to minimize a certain loss function (\eg,~mean square error (MSE) in regression tasks). Splitting terminates when nodes only contain homogeneous features or the number of observations falls below a minimum size. For prediction, a new $X$ traverses each tree’s split path to a leaf, and the RF outputs average predictions across all $B$ trees.

\subsection{A delayed acceptance sequential Monte Carlo algorithm}
\label{sec: dasmc}
Sequential Monte Carlo approaches are alternatives to MCMC for Bayesian phylogenetic inference. There are several advantages of using SMC over MCMC. First of all, we can construct a series of intermediate targets to gradually approximate the posterior, which is especially effective in complex and multimodal posterior distributions. Secondly, the parallelization of SMC methods is straightforward, which accelerates computing and predicting efficiency. Thirdly, SMC methods do not require convergence tests as consistency property holds under mild conditions, while convergence diagnosis of non-Euclidean tree spaces remains a challenge for MCMC methods. Finally, SMC can provide the estimation of marginal likelihood function as a by-product of the algorithm, contributing to Bayesian model selection.

\cite{wang2018annealed} developed the annealed sequential Monte Carlo (ASMC) algorithm for Bayesian phylogenetic inference. In the ASMC algorithm, a powered posterior series $\gamma_r(x) \propto p(x)p(D|x)^{\phi_r}$ is introduced to facilitate the exploration of complex phylogenetic tree posterior. At each SMC iteration, we iterate between propagation, weighting and resampling to achieve the approximation of the next intermediate posterior target. 
Phylogenetic MCMC moves are used in the propagation step to propose samples for the next target distribution. 
However, the total computational complexity is a linear function of the product of number of particles and intermediate target distributions. A large number of particles and intermediate targets are often required to approximate the posterior of a tree with many taxa.

In this article, we propose a more efficient delayed acceptance sequential Monte Carlo (DA-SMC). A series of intermediate target distributions are introduced
\begin{equation}
\label{eq: 6}
    \pi_r(x) \propto p(x)[\mathbf{1}(\Delta_{x}>\kappa)\min\{p(D|x), f(x)\}]^{\phi_r}\rho(x)^{1-\phi_r}.
\end{equation}
Here $\phi_r$ $(r = 1, 2, \ldots, R)$ is an increasing powered sequence with $\phi_1 = 0$ and $\phi_R = 1$, $\rho(x)$ is a reference distribution that facilitates the posterior inference. In Bayesian phylogenetic inference, $\rho(x)$ includes the distribution of evolutionary parameter $\theta$, and the distribution of the phylogenetic tree $t$. For the evolutionary parameter $\theta$, we assume the reference distribution follows the same distribution family as its prior distribution, and the parameters in the reference distribution are estimated using the training data collected. 
For example, if we use a Kimura 2-parameter (K2P) model \citep{kimura1980simple}, where the only evolutionary parameter is the transition-transversion ratio. Then we assume the ratio follows a Gamma distribution, and estimate the shape and scale parameter of this Gamma distribution using the training data. We refer readers to  Section \ref{sec: train} for detailed description of collecting training data. 
The phylogenetic tree space is combinatorial. It is not straightforward to design a parametric distribution for the tree. Here we propose the following reference distribution
$$\rho(t)\propto \frac{1}{\big(d_{PM}(t)+1\big)^2},$$ 
where $d_{PM}(t)$ represents the Partition Metric between tree $t$ and the guide tree, and the guide tree is constructed using our training data. In our construction, we assign higher weights to a tree that is closer to the guide tree.

\begin{figure}[ht]
    \centering
    \includegraphics[width=1\linewidth]{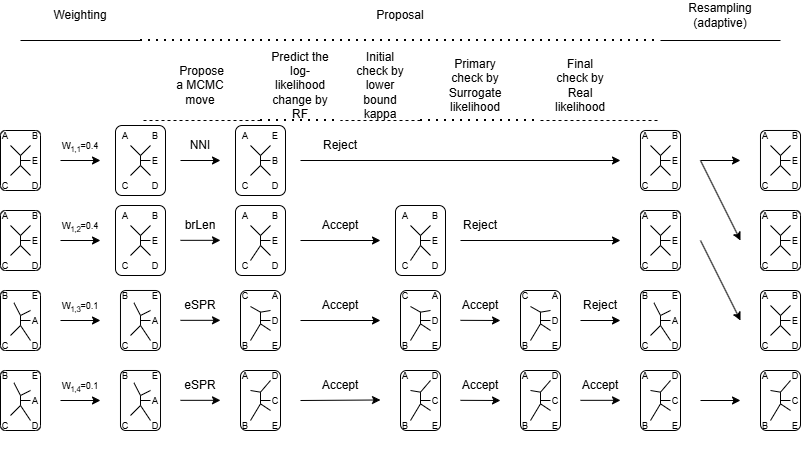}
    \caption{An overview of the Bayesian phylogenetic DA-SMC framework. Given a list of weighted particles, the DA-SMC algorithm performs the following three steps to approximate $\pi_{r+1}$: (1) In the weighting step, the weights of all particles are incrementally updated and used to evaluate the marginal likelihood; (2) In the proposal step, we first propose a new tree and record the corresponding features of the move. Then the features are sent into the random forest model for predicting the log likelihood change. Finally, we reject some proposals according to the three delayed acceptance steps in Section \ref{sec: dasmc}, and only calculate the true likelihood in the last delayed acceptance step; (3) In some iterations (\eg,~the effective sample size is below a threshold), a resampling step is applied to prune particles with small weights. }
    \label{fig:Fig2}
\end{figure}
The DA-SMC algorithm is processed for $R$ iterations and maintains a collection of latent states of size $K$, each state $\{x_{r,k}\}$ with a corresponding non-negative weight $w_{r,k}$ is called a particle.
A particle population at iteration $r$, $(x_{r,\cdot},w_{r,\cdot})=\{(x_{r,k},w_{r,k}):k\in[1,...,K]\}$, can be used to estimate posterior probabilities using the approximation 
$$\int\pi_r(x)\psi(x)dx\approx \sum_{k=1}^K W_{r,k}\psi(x_{r,k}),$$ where $W_{r,k}$ represents the normalized weight $w_{r,k}/\sum_{k'}w_{r,k'}$, and 
$\psi(x)$ is a test function. The expectation of $\psi(x)$ is approximated by the weighted particles. In the case that $\psi(x)=\mathbf{1}_{\tilde{x}_{R,k}}(x)$ is an indicator function, the expectation $\sum_{k=1}^K W_{r,k}\psi(x_{r,k})$ is an approximation of the posterior distribution $\pi(x)$.

We initialize the DA-SMC algorithm with $K$ particles. 
Given a list of weighted particles of the $r$-th intermediate target $\pi_r(x)$, the DA-SMC algorithm performs the following three
steps to approximate $\pi_{r+1}(x)$:  propagation, re-weighting and resampling. Figure \ref{fig:Fig2} provides an overview of the three steps.

{\bf Propagation:} We propagate new samples $\{x_{r+1,k}\}_{k = 1}^K$ via $\pi_{r+1}$-invariant DA-MCMC moves introduced in Section \ref{sec: daphylo}.

{\bf Re-weighting:} We compute the weight function for particles at iteration $r$ with
\begin{equation}
  W_{r+1,k}\propto w_{r,k} \cdot 
  \bigg(\frac{\mathbf{1}(\Delta_{x_{r,k}}>\kappa)\min\{p(D|x_{r,k}), f(x_{r,k})\}}{\rho(x_{r,k})}\bigg)^{\phi_{r+1} - \phi_r}.
\end{equation}
The derivation of the weight update function is shown in the Appendix S.1. Note that the weight update function $W_{r+1,k}$ only depends on particles at the $r$-th iteration. This allows us to swap the {\bf Propagation} and {\bf Re-weighting} steps, and select $\phi_r$ by controlling the degradation of the particle population when moving from $\pi_{r}$ to $\pi_{r+1}$ by targeting a specific reduction in the relative conditional effective sample size (rCESS, \citealp{zhou2016toward}), which takes the form \[\text{rCESS}(W_{r,\cdot}, \phi)=\frac{(\sum_{k=1}^K W_{r,k}p(D|x_{r,k})^{(\phi-\phi_{r})})^2}{\sum_{k=1}^KW_{r,k}p(D|x_{r,k})^{2(\phi-\phi_{r})}}.\]
For equation $\text{rCESS}(W_{r,\cdot}, \phi) = \alpha$, we use the bisection method to solve for $\phi$. Here $\alpha$ is usually a value close to $1$ in the phylogenetics context (\eg,~0.999).

{\bf Resampling:} Finally, we conduct a resampling step to prune particles with small weights. We resample $K$ particles from the empirical distribution $\hat{\pi}_{r+1}(x)\approx \sum_{k=1}^K W_{r+1,k}\delta_{x_{r+1,k}(x)}$.
 A list of equally weighted samples is obtained
after conducting the resampling step. We denote the particles after resampling by $\{\tilde{x}_{r,k}\}$. 
As recommended in the past literature \citep{wang2018annealed, 10.1093/bioinformatics/btaa867}, we only perform the resampling step when the particle degeneracy, measured by relative effective sample size (rESS), falls below a pre-determined threshold value. The relative effective sample size (rESS) is defined as 
\[\text{rESS}(W_{r,.})=\frac{1}{K\sum_{k=1}^K W_{r,k}^2},\]
which takes value in $[1/K, 1]$.

The Algorithm \ref{algo:Our DA} provides detailed description of our proposed DA-SMC algorithm. 

\begin{algorithm}[htp]
\SetAlgoLined
\SetKwFunction{IL}{InitializeDistance}
\SetKwFunction{PL}{PropagateInsertion}
\SetKwFunction{MIN}{Min}
\SetKwFunction{MX}{Max}
\SetKwFunction{TOP}{Top}
\SetKwFunction{Push}{Push}
\SetKwFunction{Pop}{Pop}
\SetKwFunction{Append}{Append}
\SetKwData{Queue}{Queue}

\KwIn{Prior $p(x)$, likelihood $p(D|x)$, surrogate abandoned lower bound $\kappa$, RF tuning constant $\delta$.}
\KwOut{Approximation $\hat{Z}$ of the marginal data likelihood $Z=p(D)$; Approximation of the posterior distribution $\sum_{k}\tilde{W}_{r,k}\delta_{\tilde{x}_{r,k}}(\cdot)\approx \pi(\cdot)$}
\textit{Initialize} SMC iteration index: $r\xleftarrow{}0$.\\
\textit{Initialize} annealing parameter: $\phi_r\xleftarrow{}0$.\\
\textit{Initialize} marginal likelihood estimate: $\hat{Z}\xleftarrow{}1$.
\For{$k\in \{1,2,\cdots,K\}$} {
    \textit{Initialize} particles: $x_{0,k}\xleftarrow{}(\theta_{0,k},t_{0,k})\sim p(\cdot)$\\
    \textit{Initialize} weights to unity: $w_{0,k}\xleftarrow{}1$
}
\For{$r\in \{1,2,\cdots,R\}$} {
   \textit{Determine} next annealing parameter:$\phi_r=F(x_{r-1,\cdot},w_{r-1,\cdot},\phi_{r-1})$.\;
   \For{$k\in \{1,2,\cdots,K\}$} {
        \textit{Compute} unnormalized weights: $w_{r, k}$.\;
        \textit{Propose} new samples $x_{r,k}^*$ (\eg,~eSPR, NNI, or branch length move).
    }
    \textit{Collect} the $K$ sets of features, and obtain  $\Delta_{x^*_{r,\cdot}}$ by RF prediction\;
    \For{$k\in \{1,2,\cdots,K\}$} {
        \uIf {$\Delta_{x_{r,k}^*}\geq\kappa$}{
            $\Delta_{x_{r,k}^*}=\Delta_{x_{r,k}^*}+\delta$ , $f(x_{r,k}^*)=\Delta_{x_{r,k}^*}+p(D|x_{r-1,k})^{\phi_{r}}$\;
            \textit{Compute} the first acceptance ratio $\tilde{\alpha}$\;
            \textit{Simulate} $u\sim \mathbf{U}(0,1)$\;
            \uIf {$u\leq\tilde{a}$}{
                \textit{Compute} the second acceptance ratio $\bar{\alpha}(x, x^\star)$\;
                \uIf {$u\leq\bar{a}$}{
                    \textit{Accept} $x_{r,k}^*$.
                }
            }
        }
        \uIf { $x_{r,k}^*$ is accepted}{
            $\tilde{x}_{r,k}=x_{r,k}^*$
        }\Else{
            $\tilde{x}_{r,k}=x_{r-1,k}$
        }
    }
    \uIf{particle degeneracy is too severe, i.e. $rESS(\tilde{W}_{r,\cdot})<\epsilon$}{    
        \textit{Update} marginal likelihood estimate $\hat{Z}$\;
        \textit{Resample} the particles.\;
        \textit{Reset} $w_{r,\cdot}=1$.
    } \Else{
        $w_{r,\cdot}=\tilde{w}_{r,\cdot}$;
        $x_{r,\cdot}=\tilde{x}_{r,\cdot}$.\;
        \textit{No} resampling is needed.
    }
}
\KwRet{All particle population $x_{r,\cdot},W_{r,\cdot}$, marginal likelihood estimate $\hat{Z}$}\;
\caption{A delayed acceptance SMC algorithm}
\label{algo:Our DA}
\end{algorithm}

Here we discuss some properties of our DA-SMC algorithm. As our SMC algorithm builds in the general framework of \cite{del2006sequential}, we generate the consistency and asymptotic normality properties from \cite{del2006sequential}. 

\noindent\textbf{Proposition 2:} Assume there is a constant $C$ such that $|\psi| \le C$ and $w_{r,k} \le C$ almost surely.  For a fixed $\phi_{r}$ $(r = 1,\ldots, R)$, the DA-SMC algorithm provides asymptotically consistent estimates:
\[
 \sum_{k=1}^K W_{r,k} \psi(x_{r,k})  \to \int \pi_r(x) \psi(x) d x ~~\text{as}~~K\to \infty,
\]
where the convergence holds in the $L^{2}$ norm sense, and $\psi$ is a test function under mild conditions (\eg,~a bounded function).

\noindent\textbf{Proposition 3:} Under the integrability conditions given in Theorem $1$ of \cite{chopin2004central}, or \cite{DelMoral2004}, section $9.4$, pages $300-306$,
\[
K^{1/2}\bigg[ \sum_{k=1}^K W_{r,k} \psi(x_{r,k})  - \int \pi_r(x) \psi(x) d x\bigg]\to N(0, \sigma_{r}^{2}(\psi)) ~~\text{as}~~K\to \infty,
\]
where the convergence is in distribution. The form of asymptotic variance $\sigma_{r}^{2}(\psi)$ depends on the resampling scheme, the Markov kernel $K_{r}$ and the artificial backward kernel $L_{r}$. We refer readers to \cite{del2006sequential} for details of this asymptotic variance.

Another well-known result in the literature is that SMC can provide an unbiased estimate of the marginal likelihood. In our DA-SMC algorithm, with a fixed sequence of $\phi_r$ $(r = 1, 2, \ldots, R)$, the marginal likelihood estimator calculated using the unnormalized incremental weights of each step of the SMC algorithm admits the following form $\hat{Z}_{R, K} = \prod_{r = 1}^R \frac{1}{K}\sum_{k = 1}^K w_{r, k}$. 

\noindent\textbf{Proposition 4:} For a fixed sequence of $\phi_r$ $(r = 1, 2, \ldots, R)$, the DA-SMC algorithm provides an unbiased estimator of marginal likelihood $Z = \int \mathbf{1}(\Delta_{x}>\kappa)p(x)\min\{p(D|x), f(x)\}dx$,
$$\mathbb{E}(\hat{Z}_{R, K}) = Z.$$

Firstly, to obtain an unbiased estimator $\hat{Z}_{R, K}$, this sequence of $\phi_r$ $(r = 1, 2, \ldots, R)$ must be prespecified before running the DA-SMC algorithm. Adaptively determining the sequence $\phi_r$ will lead to biased estimates of marginal likelihood \citep{DelMoral2004}. Secondly, due to the construction of the DA kernel, this marginal likelihood estimator differs from the true value. However, this discrepancy can be reduced by properly selecting features for the random forest model and by adjusting the threshold value. In our numerical experiments, we demonstrate that the marginal likelihood estimates provided by our DA-SMC can be used to calculate Bayes factors and effectively identify the true data-generating evolutionary model.

\subsection{Training data collection and selection of $\delta$}
\label{sec: train}
The accuracy of the predicted change of log-likelihood is critical to the quality of the inferred posterior distribution. We need sufficient training data to train the random forest. One simple strategy for collecting training data is to sample phylogenetics trees from the prior distribution, which is fast and simple to implement. Due to the complex phylogenetic tree space, most samples from the prior distribution may belong to low-posterior regions. However, 
enough training points in the high posterior region are required to ensure the accuracy of the machine learning model.

In this article, we propose to collect a training data set by using a pilot run of the annealed sequential Monte Carlo \citep{wang2018annealed}, which sequentially generates training data, with a relatively small number of particles $(K)$ and intermediate target distributions $(R)$. The main advantage of this method is that we are able to obtain samples across the entire region. Note that in addition to samples from the high posterior region, we also require samples from the low posterior region. The samples from the first few intermediate target distributions are relatively dispersed, mainly on the tails of the posterior. Through MCMC moves and resampling, they gradually move to high-posterior regions. 

In the propagation step of ASMC, we use MCMC moves with eSPR, stNNI, Multiplier and global Multiplier proposals. The reason for using these four moves is that we train the surrogate log-likelihood change caused by these four moves in random forests. Note that we record the log-likelihood change for both accepted and rejected moves, rather than only accept samples. Those rejected samples also evaluate the change in the likelihood function caused by one specific proposal, and provide information of the move. After running the ASMC, we obtain training samples across the posterior region.  
The samples of the last few intermediate target distributions can also be used to train the reference distribution shown in Eq.~(\ref{eq: 6}). 
 
The selection of delay-acceptance parameter $\delta$ is a trade-off between posterior accuracy and computing speed. This is also demonstrated in the numerical experiments of Section \ref{sec: deltasensitivity}. A proper $\delta$ is crucial for our algorithm. Here we propose a procedure for selecting $\delta$. The basic idea is to control the false rejection rate of MH moves by using the surrogate likelihood function. We describe this procedure in more detail. First, during the training data collection procedure, we record the details of each move, including the parameters for calculating acceptance probability (\eg,~prior ratio, likelihood ratio, proposal ratio, temperature) and the uniform random number. Second, after training the random forest, we predict the surrogate change for every training sample. Third, we write the acceptance probability based on the surrogate likelihood change provided by random forest as a function of $\delta$. As we have recorded all information to judge whether a sample is accepted, we can compute the false rejection rate as a function of $\delta$. This is a monotonically increasing function. Hence, with a fixed false rejection rate (\eg,~5\%), we can use the bisection method to solve for this $\delta$. 

\section{Simulation Studies}
\label{sec:sim}
\subsection{Setup}
\label{sec:setup}
To evaluate the performance of our proposed method, we conducted simulation studies under controlled conditions. 
In most cases in this section, we adopt the following simulation settings: We simulated a single random unrooted binary tree with $30$ taxa. The tree topology was sampled from a uniform distribution over all possible unrooted tree topologies and branch lengths were generated from an exponential distribution with rate parameter $10.0$. 

For the simulated ground truth tree, we generated $30$ datasets with DNA sequence alignment.
Sequences of length $2,000$ base pairs were generated for each taxon, resulting in $30$ taxa $\times$ $2,000$ bp alignments for each dataset.
We simulated the DNA sequence evolution using the K2P (Kimura 2-parameter) model (Kimura, 1980) and set the transition/transversion ratio $\kappa=2$. A K2P model indicates that the nucleotides $A,T,C,G$ have equal stationary frequencies, and all sites evolved under homogeneous rates. Additionally, sequences were simulated without missing data or gaps. 

Simulation data were generated using functions from the \emph{p4} library \citep{10.1080/10635150490445779}. In our subsequent workflow, we also made use of the MCMC objects and proposal operators provided by \emph{p4}. This foundation allowed us to implement specific modifications and extensions to the original library. In particular, we activated the eSPR move (which was originally present but not fully functional) and additionally implemented the stNNI move. To bridge the gap between the proposal mechanisms and our machine‑learning framework, we added support for feature extraction from both the eSPR and stNNI moves. 

\subsection{Tree distance evaluation}
Topological accuracy between the reference tree and the inferred tree is a vital indicator of our method. For each simulated dataset, we applied our delay-acceptance annealed SMC method to obtain weighted samples from the phylogenetic posterior distribution. Tree samples are summarized using the majority-rule consensus tree approach \citep{felsenstein1981evolutionary}, which means that a split is admitted in the consensus tree only if the split appears in over $50\%$ tree samples. The \emph{sumt} library \citep{Pedersen_sumt_a_command-line_2023} was employed to obtain majority‑rule consensus trees. 

To quantify phylogenetic accuracy, we employed some complementary evaluation approaches, such as the Robinson-Foulds (RF) metric with branch lengths \citep{RobinsonFoulds1979} and the Partition Metric (PM,  \citealp{Robinson1981}). These metrics are implemented in the Python library \emph{Dendropy} \citep{2010DendroPy}. We also measured the log-likelihood difference between the consensus tree (or the maximum-likelihood tree among the posterior samples) and the ground truth tree to assess the quality of tree estimation.

\subsection{Estimate assessment and model selection using marginal likelihood}

In this section, we conducted a systematic comparison of marginal likelihood estimates for our proposed DA-SMC. The primary objective is to evaluate the performance of DA-SMC in estimating marginal likelihoods for phylogenetic model selection. To evaluate the model selection capability of our delay-acceptance annealed SMC method, we designed a comparative analysis framework assessing marginal likelihood estimates across three standard nucleotide substitution models of increasing complexity: JC69 \citep{jukes1969evolution}, K2P \citep{kimura1980simple} and GTR \citep{rodriguez1990general}. This experimental design aims to systematically evaluate the ability of our DA-SMC to correctly identify the true data-generating model (\ie,~K2P as the true model in simulations), and to characterize performance under model misspecification scenarios.

We simulated $5$ distinct trees with $30$ taxa, and for each tree we simulated $30$ different alignments. The setup is described in Section \ref{sec:setup}. All alignments are simulated under the K2P model, yielding $150$ simulated sequences in total. 
For each simulated datasets, we adopted a DA-SMC with number of particles $K=300$ and total iterations $R=20000$ to estimate the marginal likelihood under each candidate model. The model comparison framework focuses on the log marginal likelihood estimates for Bayesian model selection.

\begin{figure}
    \centering
    \includegraphics[width=0.8\linewidth]{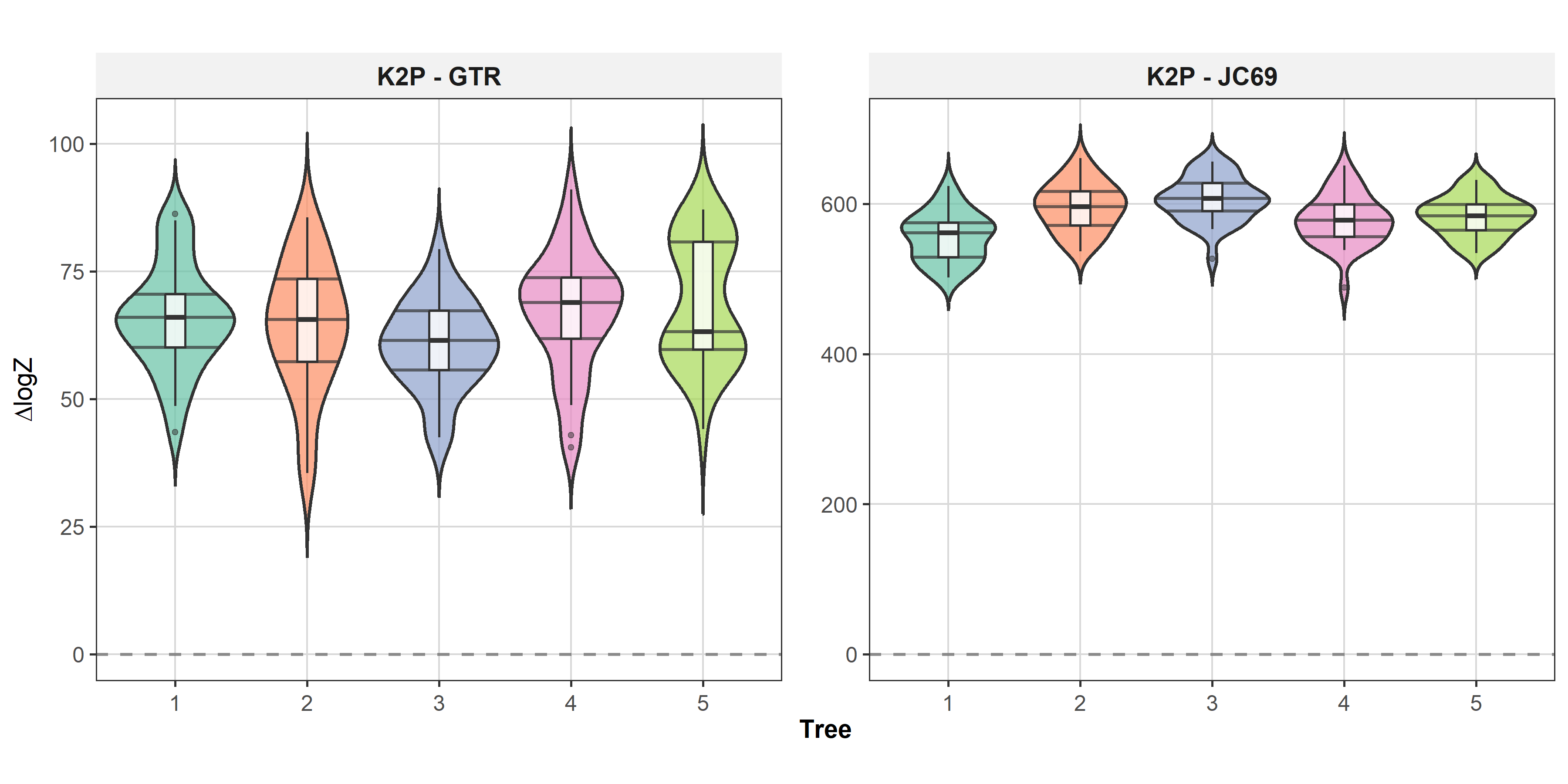}
    \caption{The logarithm marginal likelihood ($\log{Z}$) difference between K2P and GTR, K2P and JC69.}
    \label{fig:logZ}
\end{figure}

Figure \ref{fig:logZ} displays the comparative analysis of marginal likelihood estimates ($\log{Z}$) visualized through violin plots comparing pairwise difference in $ \log{Z}$ between K2P and JC69 (\ie,~K2P - JC69), and between K2P and GTR (\ie,~K2P - GTR). Our DA-SMC method demonstrates robust capability in correctly identifying the true data-generating model (K2P).
The JC69 model assumes equal transition and transversion rates, thus represents a mis-specified model that fails to capture the biologically realistic rate variation present in the simulated data. All JC69 models yield significantly lower marginal likelihood estimates under our DA-SMC.
Conversely, while the GTR model encompasses the K2P model as a special case (when appropriately parameterized), its additional parameters introduce increased model complexity without corresponding improvement in fit for these particular datasets. The resulting Bayesian penalty for over-parameterization leads to generally lower marginal likelihood estimates for GTR models. All GTR models have lower marginal likelihood estimates similarly.

These findings validate the effectiveness of our DA-SMC in Bayesian model selection, demonstrating its ability to appropriately penalize both underfitting (JC69) and over-parameterization (GTR) while correctly identifying the true model that balances biological realism with parametric parsimony.

\subsection{Sensitivity analysis of SMC parameter $N,K,R$ }
To comprehensively evaluate the performance of our DA-SMC algorithm in terms of phylogenetic tree inference, we conducted a systematic parameter sensitivity analysis. This investigation focuses on SMC parameters: the number of particles of SMC $K$, the SMC iterations $R$ which is controlled by $\beta=1-\alpha$ in DA-SMC, and the overall number of samples generated in the SMC $N$ ($N=K\times R$). 
We conducted three sets of sensitivity analyses to investigate the effects of varying $R$ and $K$ respectively.

\begin{figure}
    \centering
    \includegraphics[width=0.95\linewidth]{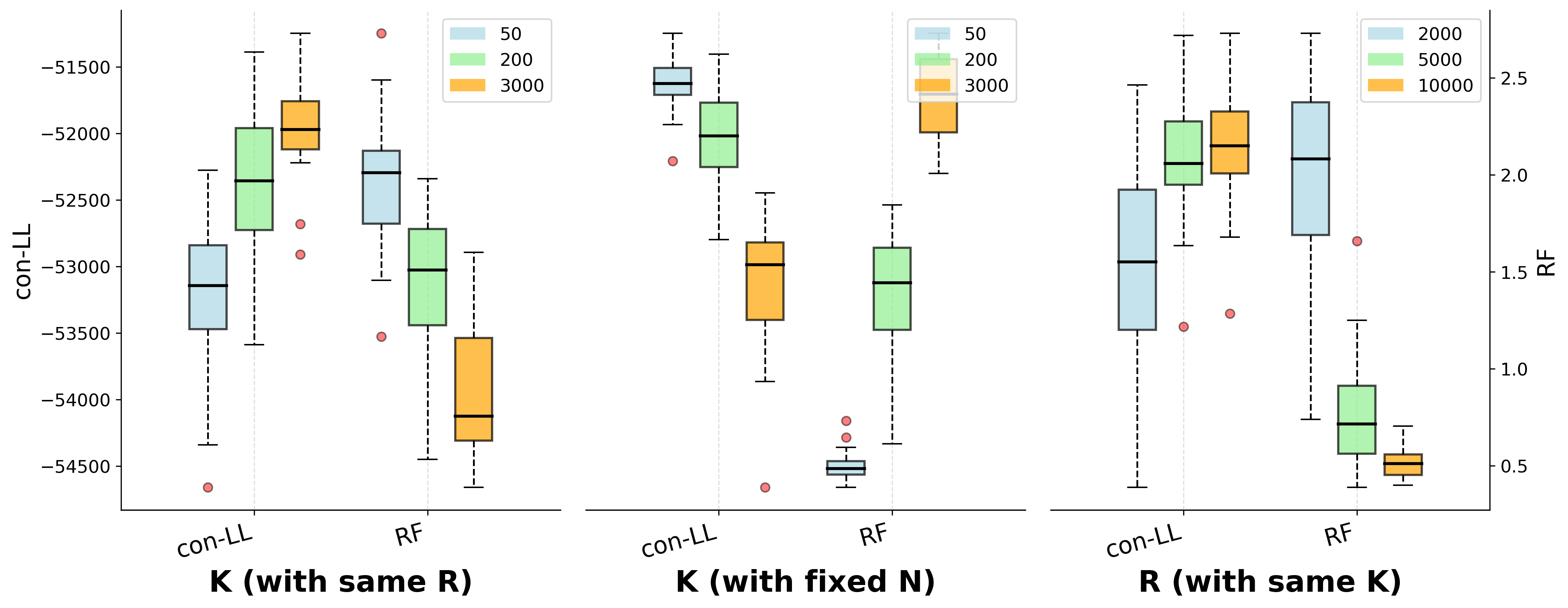}
    \caption{Consensus log-likelihood and RF distance as a function of $K$ (with fixed $R$), $K$ (with fixed $N$) and $R$ (with fixed $K$). Left: $K$ (with fixed $R$); Middle: $K$ (with fixed $N$); Right: $R$ (with fixed $K$). See more metrics in Figure \ref{fig:figNKR_supp} in Supplementary Appendix S.4.}
    \label{fig:figNKR}
\end{figure}

Firstly, the left and right sub-figures in Figure~\ref{fig:figNKR} show that when one of $K$ and $R$ is held constant, increasing the other will consistently improve SMC convergence, as evidenced by both higher log-likelihood values and greater similarity to Moreover, the middle sub-figure in Figure~\ref{fig:figNKR} shows that under a fixed total sampling budget ($N$), longer iteration schemes \(R\) are favored. These findings support the theoretical principle that, within reasonable bounds, iteration depth contributes more critically to SMC convergence than particle quantity. On the other hand, given the computational efficiency of the prediction step (whose runtime remains largely invariant to particle count), employing larger particle sets may serve as a viable alternative, while ensuring sufficient number of iterations maintained.

\subsection{Sensitivity analysis of delay acceptance parameter $\delta$}
\label{sec: deltasensitivity}

\begin{figure}[h!]
    \centering
    \includegraphics[width=0.8\linewidth]{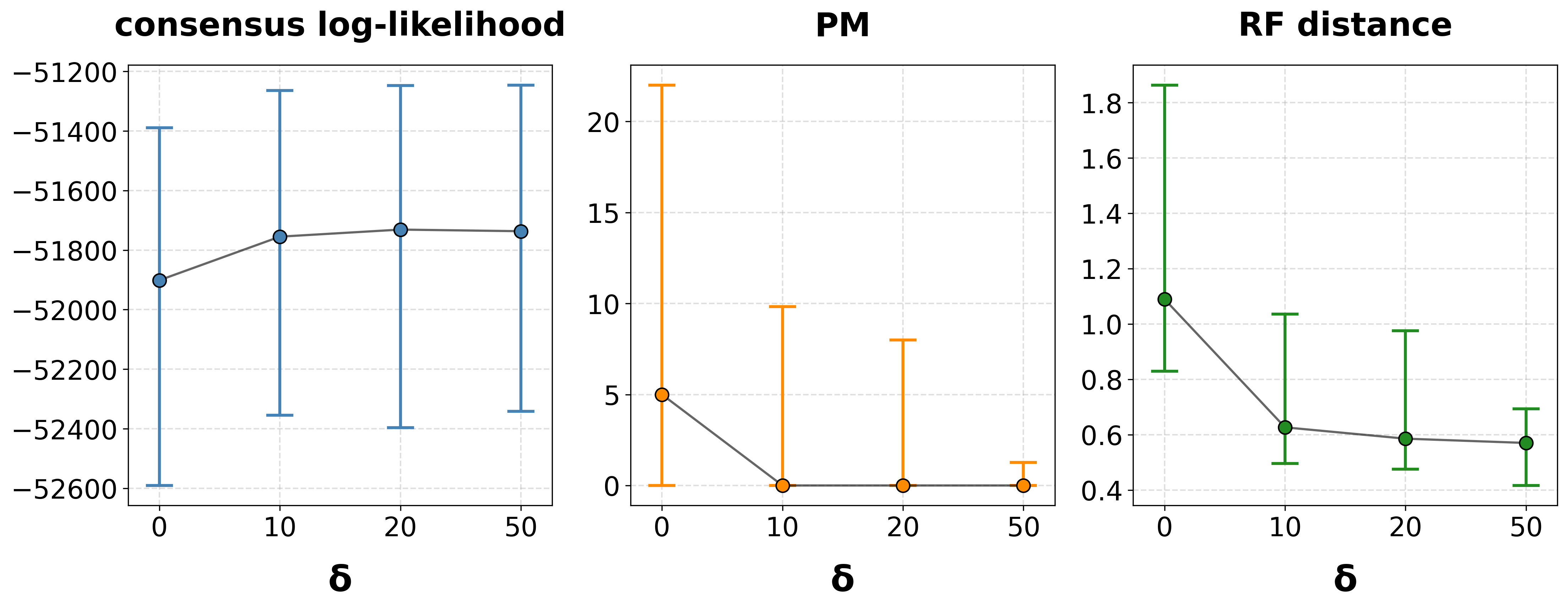}
    \caption{Consensus log-likelihood, PM and RF distance as a function of different $\delta$.}
    \label{fig:figdelta}
\end{figure}

The experimental design enables us to identify optimal delay-acceptance parameter configurations that balance computational efficiency and phylogenetic accuracy. 
The analysis of the likelihood tuning bias ($\delta$) indicates that increasing this parameter (with other SMC hyper-parameters fixed) improves the overall performance of the DA-SMC algorithm across key metrics consistently (see Figure~\ref{fig:figdelta}). However, this improvement is attained at the cost of increased computational runtime. Theoretically, an excessively large $\delta$ value allows all particles to pass the surrogate-likelihood filtering step, causing the algorithm to degenerate into a standard SMC without delayed acceptance. Empirically, on the simulated datasets of moderate complexity used in this study, increasing $\delta$ beyond about $100$ yields diminishing returns, providing no substantial further improvement in convergence stability or tree structure accuracy. 

\subsection{Timing experiments}
\label{sec: timingexperiment}
In this section, we show the time savings achieved by the DA-SMC due to the design of DA-kernel and the acceleration of computing speed achieved by parallelization property of SMC. We simulated $4$ distinct trees with $20,40,80,160$ taxa, and for each tree we simulated a set of alignments with $1000$ bases per sequence. All alignments are simulated under the K2P model. 
\subsubsection{Time savings achieved by the DA-kernel}
\label{sec: computingefficiency}

\begin{figure}
    \centering
    \includegraphics[width=0.8\linewidth]{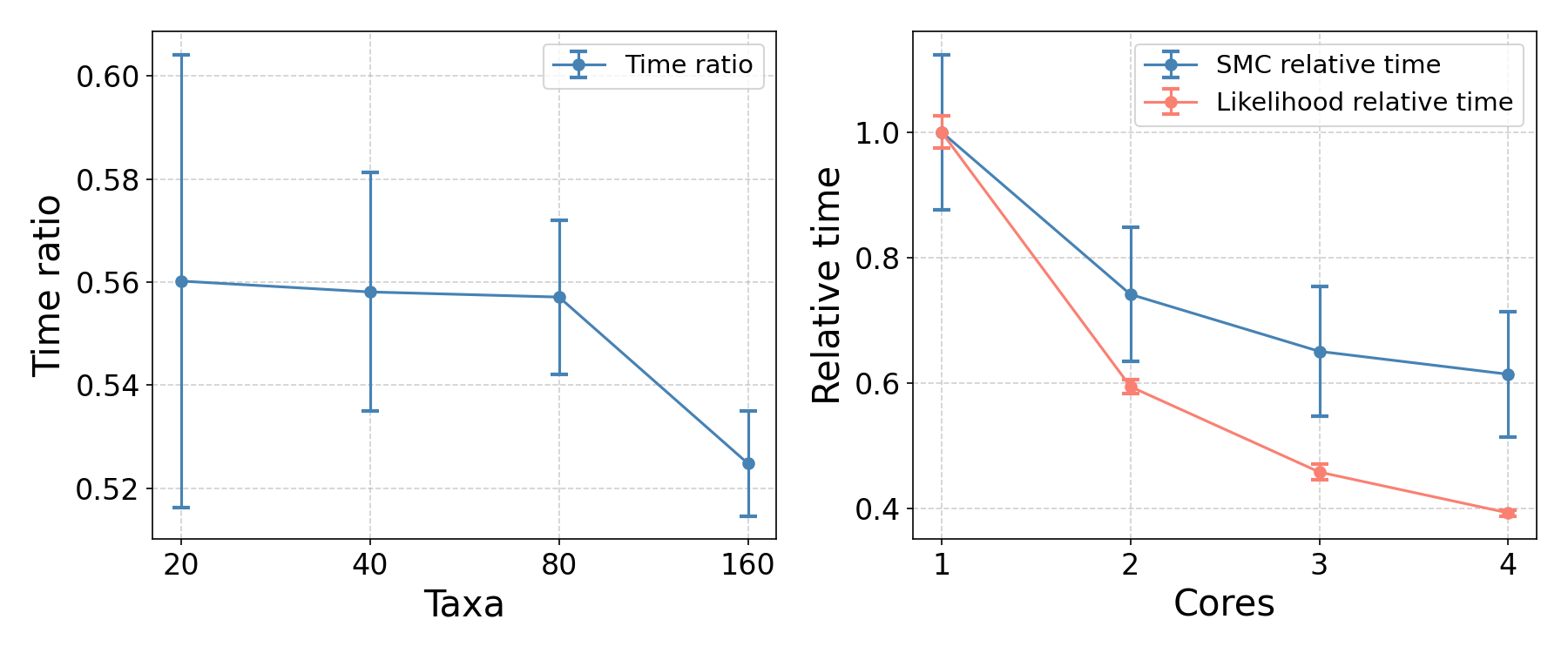}
    \caption{Left: Runtime ratio of DA-SMC to ASMC as a function of the number of taxa. A lower ratio indicates greater efficiency of DA-SMC. Right: Relative runtime of parallelized DA-SMC (likelihood evaluation) with different numbers of cores.}
    \label{fig:time}
\end{figure}

After the pilot SMC run and random forest training, for each tree we conducted $20$ repeated DA-SMC and $20$ ASMC runs with number of particles $K =1000$. The left panel of Figure \ref{fig:time} shows the time-cost ratio of our DA-SMC over the standard ASMC \citep{wang2018annealed}. The results demonstrate that the computational efficiency advantage of DA-SMC is more evident for larger phylogenetic data sets. 

\subsubsection{Parallelization} 
The propagation and weighting steps of SMC can be easily parallelized by allocating particles across different cores. This is another advantage of SMC over MCMC methods. We used the simulated data (Section \ref{sec: timingexperiment}) with $40$ taxa, and conducted $20$ repeated DA-SMC with different number of cores. 
The right panel of Figure \ref{fig:time} shows the total SMC runtime and likelihood calculation time for different numbers of cores in our parallelized version of DA-SMC, relative to the runtime using a single core. The results demonstrate that parallelization can significantly improve the computational speed of DA-SMC. Moreover, the improvement in likelihood evaluation is even more pronounced, highlighting the advantage of DA-SMC for larger datasets, where likelihood evaluation remains a bottleneck to computational efficiency.

\section{Real data analysis}
\label{sec:real}
We apply our proposed approach to three real data sets of DNA sequences from TreeBASE \citep{piel2009treebase}. The {\it primates} data set contains DNA sequences, each of length 898 for 12 taxa. {\it M336} contains DNA sequences, each of length 1949 for 27 taxa. {\it M1809} contains DNA sequences, each of length 1824 for 59 taxa. 

We used MrBayes (with the default setting) with the same computational budget as the compatible comparison. Here the computational budgets of MrBayes refer to the number of MCMC iterations, and for SMC, it refers to the number of SMC iterations multiply the number of particles.  
For comparison, we used multiple metrics including the log-likelihood of consensus tree (ConsensusLL), maximum log-likelihood of posterior samples (BestLL), Robinson-Foulds (RF) metric, Partition Metric (PM) and Log marginal likelihood estimates (LogZ). Because our DA-SMC method does not require evaluating the likelihood function at every MCMC move, the total number of likelihood evaluations in our method is lower than that in the MrBayes.
The reference trees of ground truth used to compute tree distances for dataset M336 and M1809 are provided by \cite{Lakner01022008}, which are based on at least six independent long MrBayes parallel tempering runs. For the {\it primates} dataset, the ground truth was obtained by running the MrBayes MCMC with the same computational setting as in \cite{Lakner01022008}.
Similar to the Section \ref{sec:sim}, we considered three candidate evolutionary models (JC69, K2P, GTR), and select the optimal model for each dataset by using marginal likelihood estimates provided by DA-SMC.

The surrogate model of our DA-SMC algorithm employs a random forest regressor to predict log-likelihood change caused by a phylogenetic tree move. 
To optimize the performance of random forests, we conducted a comprehensive parameter sensitivity analysis focusing on key hyper-parameters that influence both predictive accuracy and computational efficiency.
We conducted a systematic grid search experiment to evaluate the impact of random forest hyper-parameters on the overall DA-SMC performance.
We investigated some key hyper-parameters across multiple values, including Number of Trees, Maximum Tree Depth, Maximum Features per Split, etc.
Training data generation process is the same as described in Section \ref{sec: train}, with a training/test split of $90\%/10\%$, and we selected the model parameters by a grid cross-validation.
For evaluation, each parameter combination is evaluated using multiple complementary metrics, including predictive accuracy (MSE) and DA-SMC performance.

\subsection{Dataset primates}

We used $K= 500$ and $\beta = 4$ for the DA-SMC algorithm, which results in a total number of $1.2\cdot 10^6$ iterations.  
For the model selection, our DA-SMC provided logZ estimates with $-6427.90$, $-6216.01$ and $-6036.63$ for models JC69, K2P and GTR. Thus the GTR model achieves the highest marginal likelihood estimates, which is consistent with the model selection result provided by MrBayes.
With the GTR model, the RF metrics of DA-SMC are lower than MrBayes, while both PM metrics converge to 0. Our DA-SMC also provided higher log-likelihood value and log marginal likelihood (logZ) estimate than MrBayes, which indicates that our DA-SMC is more computationally efficient. 

\begin{figure}[h!]
    \centering
    \includegraphics[width=1\linewidth]{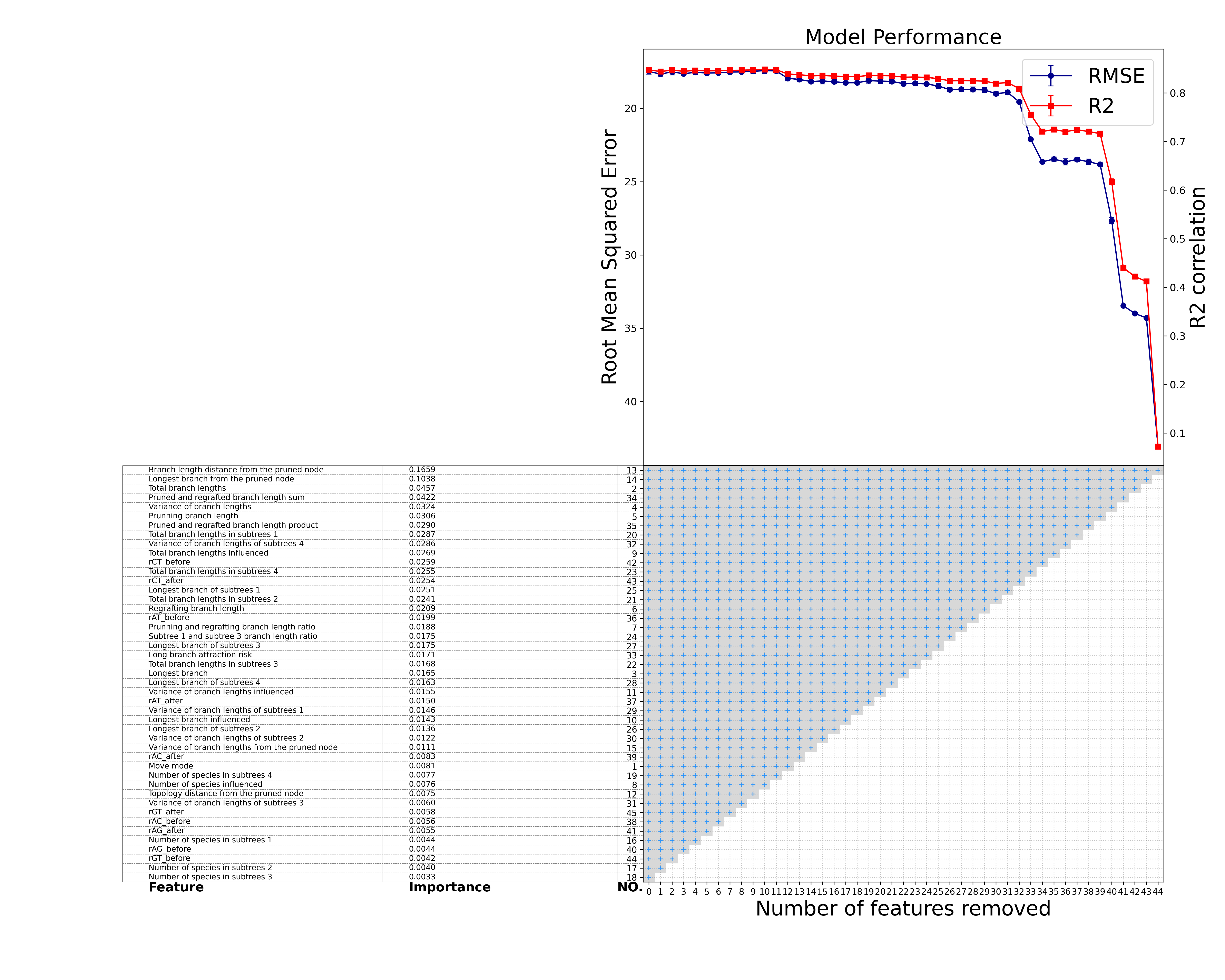}
    \caption{The RMSE (Root Mean Square Error) and $R^2$ obtained with decreasing number of features for Random Forest training. The table at the bottom displays the feature composition within each set of features along with their importance, as determined by the importance-decreasing order.}
    \label{fig:feature_remain_primates_GTR}
\end{figure}

Figure \ref{fig:feature_remain_primates_GTR} shows the feature selection of Random Forest model for dataset {\it primates}. The features in the sub-table are sorted by descending importance from the model training. We removed features stepwise from the tail (least important). The low RMSE and high $R^2$ indicate satisfactory random forest performance with the full feature set. Performance declines slowly at first as features are reduced, but after a threshold, it drops rapidly. The results suggest that at least $15$ features should be kept for relatively sufficient performance. In more complex datasets like {\it M336} or {\it M1809}, the dropping threshold occurs earlier, so more features are advised to be kept, see Supplementary Appendix S.4 for details.

\begin{table}[H]
    \centering
     \caption{Comparison of DA-SMC and MrBayes for dataset {\it primates}.}    
    \label{tab:placeholder}
    \begin{tabular}{ccc}\toprule
         Method&  DA-SMC& MrBayes\\\midrule
         ConsensusLL&  -5935.87& -5943.63\\
         BestLL&  -5938.20& /\\
         PM &  0& 0\\
         RF&  0.0302& 0.0562\\
         LogZ &  -6036.63& -6068.90\\ \bottomrule
    \end{tabular}
\end{table}

\subsection{Dataset M336}
We used $K= 500$ and $\beta = 5.3$ for the DA-SMC algorithm, which results in $8\cdot 10^6$ iterations.  
For the model selection, our DA-SMC provided logZ estimate with $-7111.98$, $-7085.58$ and $-7052.93$ for models JC69, K2P and GTR. The GTR model achieves the
highest marginal likelihood estimates, which is consistent with the model selection result provided by MrBayes.
With the GTR model, the RF metrics of DA-SMC are lower than MrBayes, and the PM metric of DA-SMC converges to 0 while MrBayes fails. Our DA-SMC also provided higher log-likelihood value and log marginal likelihood (logZ) estimate than MrBayes, which indicates that our DA-SMC is more computationally efficient.

\begin{figure}
    \centering
    \includegraphics[width=1\linewidth]{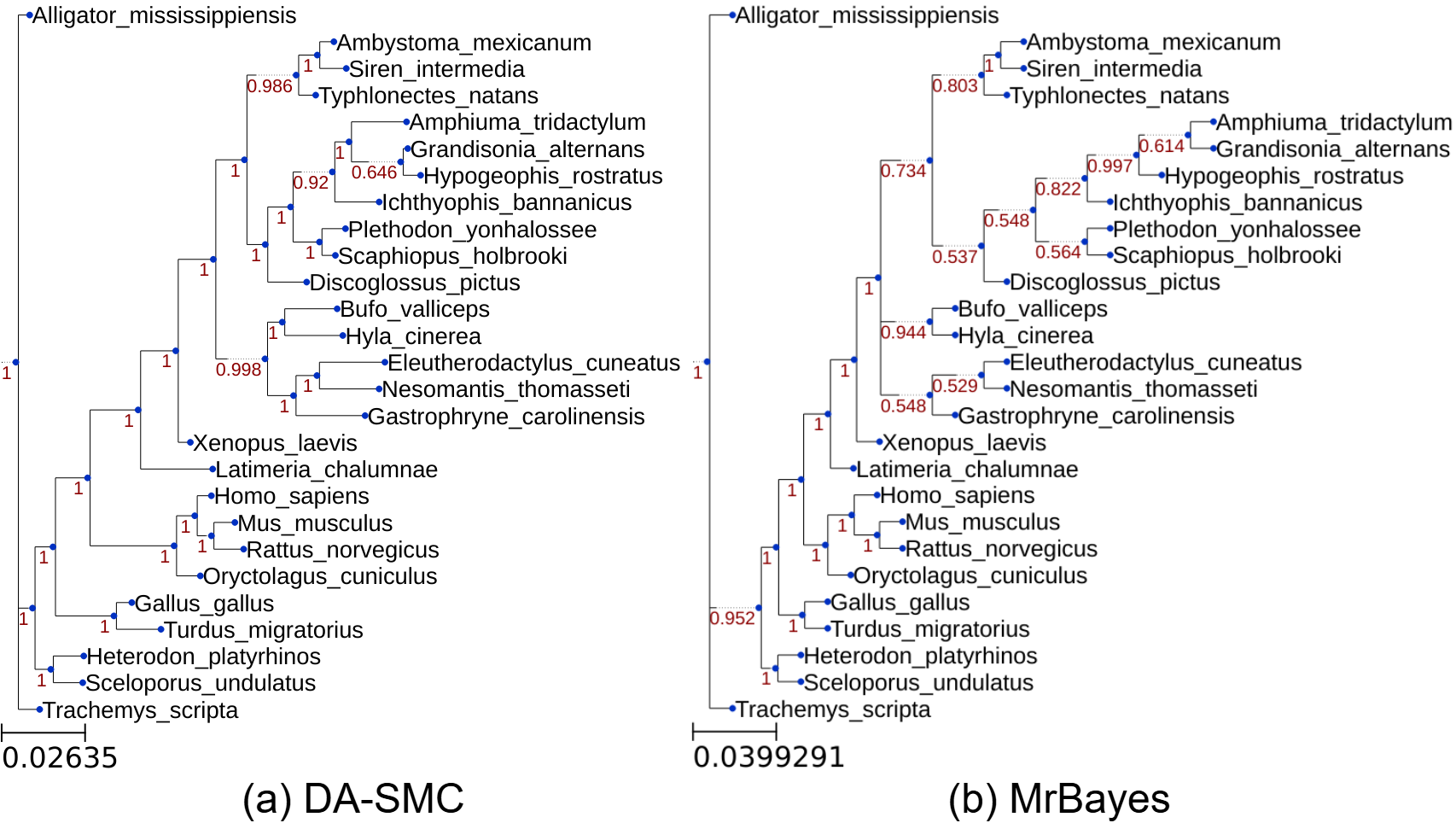}
    \caption{The majority-rule consensus trees for the M336 dataset estimated by (a) DA-SMC and (b) MrBayes. The numbers on the splits represent the clade posterior probabilities.}
    \label{fig:consensus_M336}
\end{figure}

Figure \ref{fig:consensus_M336} shows the majority-rule consensus trees provided by DA-SMC and MrBayes for dataset {\it M336}. The clade posterior probabilities are shown on the corresponding splits. The clade posterior probabilities further demonstrate that DA-SMC provided high confidence, with most clades supported by probabilities exceeding $0.99$. In contrast, MrBayes yields less convincing results, with lower clade probabilities. Thus, DA-SMC exhibited superior performance in estimating tree posteriors. 

\begin{table}[H]
    \centering
     \caption{Comparison of DA-SMC and MrBayes for dataset M336.}    
    \label{tab:Comparison_M336}
    \begin{tabular}{ccc}\toprule
         Method&  DA-SMC& MrBayes\\\midrule
         ConsensusLL&  -6777.11& -6777.25\\
         BestLL&  -6781.95& /\\
         PM &  0& 2\\
         RF &  0.0875& 0.2157\\
         LogZ &  -7052.93& -7074.15\\ \bottomrule
    \end{tabular}
\end{table}

\subsection{Dataset M1809}
We used $K= 1000$ and $\beta = 5$ for the DA-SMC algorithm, which results in $2\cdot 10^7$ iterations.  
For the model selection, our DA-SMC provided logZ estimate with $-37554.87$, $-35613.14$ and $-34427.81$ for models JC69, K2P and GTR. The GTR model achieves the highest marginal likelihood estimates, which is consistent with the model selection result provided by MrBayes.

Table \ref{tab:Comparison_M1809} shows that, under the GTR model, the RF and PM metrics of DA-SMC were lower than those of MrBayes. DA-SMC also achieved a log-likelihood value comparable to that of MrBayes. The posterior means and credible intervals (CIs) of the evolutionary parameters were reported in Table \ref{tab: posteriorM1809}. In addition, we controlled the false-negative rate at $10\%$ by tuning $\delta$ and monitored the proportion of particles that bypassed the exact likelihood calculation. This proportion reached $46.64\%$, resulting in a substantial reduction in the number of expensive likelihood evaluations.

\begin{table}[H]
    \centering
      \caption{The posterior mean and credible intervals (CI) of evolutionary parameters for dataset {\it M1809}.}
          \label{tab: posteriorM1809}
    \begin{tabular}{c|ccc}
     \hline
         Parameter&  $\pi(A)$&  $\pi(T)$& $\pi(C)$\\
        
         Mean (CI)&  0.2224 (0.2101, 0.2323)&  0.3910 (0.3748, 0.4052)& 0.2204 (0.2087, 0.2323)\\
         \hline
         Parameter&  $\pi(G)$&  $r(A\leftrightarrow T)$& $r(A\leftrightarrow C)$\\
       
         Mean (CI)&  0.1662 (0.1539, 0.1804)&  0.0576 (0.0544, 0.0582)& 0.0576 (0.1978, 0.2147)\\
         \hline
         Parameter&  $r(A\leftrightarrow G)$&  $r(C\leftrightarrow T)$& $r(G\leftrightarrow T)$\\
      
         Mean (CI)&  0.3153 (0.2943, 0.3170)&  0.3962 (0.3928, 0.4106)& 0.0078 (0.0065, 0.0120)\\
         \hline
 Parameter& $r(C\leftrightarrow G)$& &\\
 Mean (CI)& 0.0248 (0.0226, 0.3138)& &\\
    \hline
    \end{tabular}
\end{table}

\begin{table}[H]
    \centering
     \caption{Comparison of DA-SMC and MrBayes for dataset M1809.}    
    \label{tab:Comparison_M1809}
    \begin{tabular}{ccc}\toprule
         Method&  DA-SMC& MrBayes\\\midrule
         ConsensusLL&  -33706.00& -33701.80\\
         BestLL&  -33708.42& /\\
         PM &  4& 7\\
         RF &  1.4775& 1.4925\\
         LogZ &  -34427.81& -34234.62\\ \bottomrule
    \end{tabular}
\end{table}

\section{Conclusion}
\label{sec: conclusion}
In this article, we introduce a novel sequential Monte Carlo framework for Bayesian phylogenetic inference with a delayed acceptance mechanism. 
The delayed acceptance mechanism enables early rejection of costly likelihood evaluations, substantially reducing computational overhead. Our empirical comparisons with standard SMC confirm these gains in efficiency. The tuning parameter $\delta$ controls the rejection rate based on the surrogate, ensuring that the false negative rate does not exceed a specified threshold. We further provide a data-driven procedure for selecting $\delta$ to achieve the pre‑specified false negative rate. 

Under the proposed delayed acceptance kernel, if the surrogate function matches the true likelihood perfectly, DA-SMC will recover the exact posterior. In practice, inaccuracies in the surrogate inevitably lead to both false rejections and false acceptances. The latter can distort the posterior distribution, and hence lead to biased posterior estimation. Increasing $\delta$ mitigates both types of error, at the cost of more evaluations of the true likelihood. For settings where posterior accuracy is paramount and computational efficiency can be partially sacrificed, we also introduce a modified delayed acceptance kernel that entirely eliminates false acceptances, guaranteeing asymptotically exact inference. 

In this work, we employ a random forest model as a surrogate likelihood function predictor by extracting detailed features of the original tree and the proposed move. The surrogate model predicts the resulting change in likelihood of each MCMC proposal. We choose random forest due to its robustness, efficiency and accessible feedback, though under the delayed acceptance kernel the framework is readily extensible to other machine learning or deep learning models, or any computationally efficient approximations of the likelihood. 

Our numerical experiments on both simulated and real datasets demonstrate the effectiveness of the proposed DA-SMC method. 
DA-SMC exhibits properties analogous to the underlying SMC framework: increasing either the number of iterations $R$ or the number of particles $N$ improves posterior accuracy, with the number of iterations being a more critical factor. Within a reasonable range, moderately increasing the tuning parameter $\delta$ yields noticeable gains in posterior precision, at little cost to computational efficiency. We also demonstrate that our method can correctly identify the true underlying evolutionary model, which highlights the reliability of the marginal likelihood estimates produced by DA-SMC. Finally, a comparison with a standard SMC under the same framework confirms the computational efficiency of the delayed acceptance mechanism.
On real datasets, we compare DA-SMC with MrBayes under identical sampling budgets. Across three datasets of varying complexity, DA-SMC consistently outperforms MrBayes in terms of both the Robinson–Foulds (RF) distance and the estimated marginal log-likelihood (logZ). 

There are several lines of future research for improvement. 
Firstly, the surrogate likelihood used in the delayed acceptance kernel currently relies on random forests fed with a set of hand‑crafted features derived from the proposed MCMC move. Future work could explore richer representations of tree space - such as the tree-to-vector bijection proposed by \cite{2025PhyloVAE}, coupled with more accurate and computationally efficient predictive models. Such enhancements would yield further gains in both the accuracy and efficiency of DA-SMC.

The second line of future work involves improving the proposal mechanism. While standard MCMC proposals often suffer from low acceptance rates, one could leverage the collected features to construct adaptive proposals. We explore providing generative proposals that are more likely to be accepted along the lines of \cite{10.1093/sysbio/syab004}. Naturally, such adaptive proposals must be carefully designed to ensure that the associated acceptance probabilities remain tractable and that their integration into the DA-SMC framework does not adversely affect the weight updates or other algorithmic steps.

Finally, the DA-SMC algorithm is inherently easy to parallel. 
If a prediction model supporting incremental learning can be adopted, the entire workflow could be streamlined: one could iteratively train the surrogate online using features collected from ongoing iterations, thereby consolidating all steps into a single DA-SMC run without the need for a separate pilot phase or model training.

\section*{Acknowledgements}
This project is supported by the National Natural Science Foundation of China (No. 12101333), the startup fund of ShanghaiTech University, the Institute Development Fund and the HPC Platform of ShanghaiTech University. 
We are grateful to Peter Foster for developing and maintaining the \emph{p4} library (Foster, 2004), which provided the core MCMC infrastructure and tree proposal mechanisms that we built upon.

\section*{Disclosure Statement}
The authors report there are no competing interests to declare.

\bigskip
\begin{center}
{\large\bf SUPPLEMENTAL MATERIALS}
\end{center}

\begin{description}

\item[Appendix S.1:] The proofs of theoretical results.

\item[Appendix S.2:] Description of selected features.

\item[Appendix S.3:] 
Illustration of tree moves.

\item[Appendix S.4:] 
Some numerical results not shown in the main text.

\item[Appendix S.5:] A summary of notations used in the paper.

\end{description}


\end{document}